\documentclass[preprint2]{aastex}

\shorttitle{Curvature Wavefront Sensing revisited}
\shortauthors{O. Guyon}
\usepackage{graphicx}
\usepackage[mathscr]{eucal}
\begin{document}

\title{High Sensitivity Wavefront Sensing with a non-linear Curvature Wavefront Sensor}
       
\author{Olivier Guyon}

\affil{Subaru Telescope, 650 N. A'ohoku Place, Hilo, 96720 HI, USA}
\affil{Steward Observatory, University of Arizona, 933 N Cherry Ave., Tucson, 85721 AZ, USA}
\email{guyon@naoj.org}

\begin{abstract}
A new wavefront sensing approach, derived from the successful curvature wavefront sensing concept but using a non-linear phase retrieval wavefront reconstruction scheme, is described. The non-linear curvature wavefront sensor (nlCWFS) approaches the theoretical sensitivity limit imposed by fundamental physics by taking full advantage of wavefront spatial coherence in the pupil plane. Interference speckles formed by natural starlight encode wavefront aberrations with the sensitivity set by the telescope's diffraction limit $\lambda$/D rather than the seeing limit of more conventional linear WFSs. Closed-loop adaptive optics simulations show that with a nlCWFS, a 100 nm RMS wavefront error can be reached on a 8-m telescope on a $m_V  = 13$ natural guide star.
The nlCWFS technique is best suited for high precision adaptive optics on bright natural guide stars. It is therefore an attractive technique to consider for direct imaging of exoplanets and disks around nearby stars, where achieved performance is set by wavefront control accuracy, and exquisite control of low order aberrations is essential for high contrast coronagraphic imaging. Performance gains derived from simulations are shown, and approaches for high speed reconstruction algorithms are briefly discussed.
\end{abstract}

\section{Introduction}
Adaptive Optics (AO) systems are now an essential part of ground-based telescopes, and routinely deliver diffraction limited images at near-IR and mid-IR wavelengths on 8 to 10 m telescopes. For several key science applications, the required wavefront quality is however still well beyond what current systems deliver. For example, obtaining high quality diffraction-limited images at visible wavelengths requires residual wavefront errors to be below 100 nm RMS. Direct imaging of exoplanets or circumstellar debris disks with a high contrast coronagraphic camera is even more demanding, and requires exquisite control of low order aberrations and mid-spatial frequencies. 

In the last decade, several high performance coronagraphs concepts have been developed in the laboratory, and full AO+coronagraph laboratory systems have demonstrated that both deformable mirror and coronagraph technologies are now sufficiently mature to allow direct imaging of planets down to Earth mass provided that wavefront aberrations are small: for example, the vacuum High Contrast Imaging Imaging Testbed (HCIT) a NASA JPL has achieved better than $10^{-9}$ raw PSF contrast at 4 $\lambda$/D separation \citep{trau07}. This high contrast value is orders of magnitude better than, at a few $\lambda/D$, the $\approx10^{-3}$ best raw PSF contrast currently achieved in the near-IR on 8 m to 10 m telescopes and the expected $\approx 10^{-4}$ raw PSF contrast for ÒExtreme-AOÓ systems currently under assembly such as SPHERE \citep{beuz08} on the Very Large Telescope (VLT) and Gemini Planet Imager \citep{maci08}. The large gap between laboratory demonstrations and on-sky performance is entirely due to wavefront control accuracy. A key difference between laboratory systems and on-sky conditions is the timescale of wavefront variations, which requires fast wavefront sampling and therefore fundamentally limits the number of photons available for sensing on ground-based systems.

The wavefront information which can be extracted from a given number of photons is known from first principles, and accurately defines the "ideal" performance which can be reached by AO systems on ground-based telescopes \citep{guyo05}. Direct comparison between this ideal limit and the sensitivity offered by current wavefront sensing schemes reveals a huge performance gap. Commonly used wavefront sensors such as Shack-Hartmann and Curvature are very robust and flexible, but are poorly suited to high quality wavefront measurement: for low-order aberrations on 8 to 10 m telescopes, they require $\approx$ 100 to 1000 times more photons than more optimized wavefront sensing techniques. While they offer nearly ideal sensitivity at the spatial frequency defined by the subaperture spacing, they suffer from poor sensitivity at low spatial frequencies, which are most critical for direct imaging of exoplanets and disks.

This paper describes a high sensitivity wavefront sensing approach which is optically derived from the curvature wavefront sensing technique, and which shares the same non-linear reconstruction algorithm as used for phase diversity. In phase diversity, images near the focal plane are used for wavefront reconstruction with a non-linear algorithm, while curvature wavefront sensing uses images near the pupil plane and a linear wavefront reconstruction algorithm. Phase diversity is theoretically more sensitive than linear curvature wavefront sensing \citep{fien98}, but, for ground-based adaptive optics applications, suffers in practice from smearing of focal plane speckles due to chromatic effects and time averaging of the fast-changing PSF. I propose in this paper a solution to improve curvature WFS sensitivity by using non-linear signals in the near-pupil images, and offer a solution to optically mitigate the chromatic smearing of the speckles which would otherwise reduce sensitivity. Unlike phase diversity, the proposed scheme is robust against time averaging of wavefront aberrations, which gradually brings the proposed technique closer to a less sensitive but stable linear curvature wavefront sensor.

The non-linear curvature wavefront sensor (nlCWFS) principle is described in \S\ref{sec:nlCWFS}. The wavefront reconstruction algorithm for the nlCWFS is described in \S\ref{sec:algo}. The WFS performance is evaluated through numerical simulations in \S\ref{sec:perf} and discussed in \S\ref{sec:disc}.

\section{nlCWFS Principle}

\label{sec:nlCWFS}

\begin{figure*}
\includegraphics[scale=0.37]{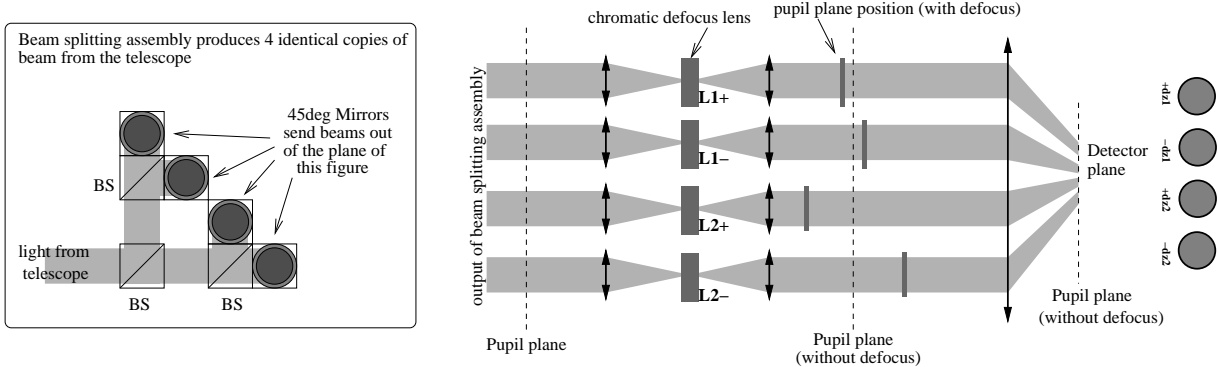}
\caption{\label{fig:nlcurvaoprinc} Conceptual layout of a non-linear Curvature wavefront sensor. See text for details.}
\end{figure*}

Figure \ref{fig:nlcurvaoprinc} shows the proposed optical implementation for the non-linear Curvature WFS (nlCWFS). In this section, the main differences from a conventional curvature WFS are explained:
\begin{itemize}
\item{The nlCWFS achieves its sensitivity by using non-linear signal in images that are acquired in planes which are considerably further away from the pupil, where non-linear effects are strong (\S\ref{ssec:largedefoc})}
\item{While curvature systems use a variable focal length vibrating membrane, here, beam splitters and lenses are used to produce the required conjugation planes on the detector(s) array (\S\ref{ssec:beamsplitters})}
\item{Chromatic lenses are used in the sensor's focal planes for compensation of chromatic effects ( \S\ref{ssec:chromatlenses} )  }
\item{Four conjugation planes are created instead of two in conventional curvature (\S\ref{ssec:nbplanes})}
\end{itemize}

\subsection{Non-linear signal and large defocus distances}
\label{ssec:largedefoc}
The performance of conventional Curvature WFS (cCWFS) is strongly constrained by the requirement that wavefront sensor signals must be a linear function of the input wavefront aberrations \citep{guyo08}. This linear constraint keeps the defocus distance much smaller than it should be for optimal conversion of phase aberrations into intensity fluctuations. The dual stroke linear scheme proposed in \cite{guyo08} only slightly mitigates the negative effects of these constraints, but linearity is still enforced.

A far more serious limitation is that linear reconstruction fully ignores the non-linear coupling between high and low order aberration in the defocused pupil images. This is best visualized by direct comparison of linear vs. non-linear measurement of tip-tilt from defocused pupil images.

In cCWFS, tip-tilt is measured as an overall displacement of the pupil in the defocused images: only the edges of the defocused pupil images are used to extract signal. The displacement is proportional to the defocus distance $dz$. If the wavefront is otherwise perfectly flat, tip-tilt is theoretically best sensed by an extremely large (infinite) defocusing distance \citep{vand02b}, where the defocused pupil image is in fact close to a focal plane image, and the tip-tilt measurement sensitivity is equivalent to what is achieved when measuring tip-tilt from a properly sampled $\lambda/D$-wide focal plane diffraction-limited image. This is hardly realistic, as higher order wavefront aberrations also need to be sensed, and the defocus distance must therefore be kept small, and not all photons should be allocated to tip-tilt sensing. Even if the defocus distance were very large, wavefront aberrations would preclude diffraction-limited imaging at the wavefront sensing wavelength. The defocused pupil plane images are therefore convolved by the $\lambda/r_0$-wide atmospheric seeing kernel (were $r_0$ is the Fried parameter quantifying the strength of the atmospheric turbulence): unless the AO correction is quite good, the tip-tilt sensitivity is at best equivalent to measuring the centroid of a $\lambda/r_0$-wide spot. It is in fact not as good as that, as the defocus distance is finite, and tip-tilt is therefore only measured from detectors at the edge of the pupil.

\begin{figure*}
\includegraphics[scale=0.34]{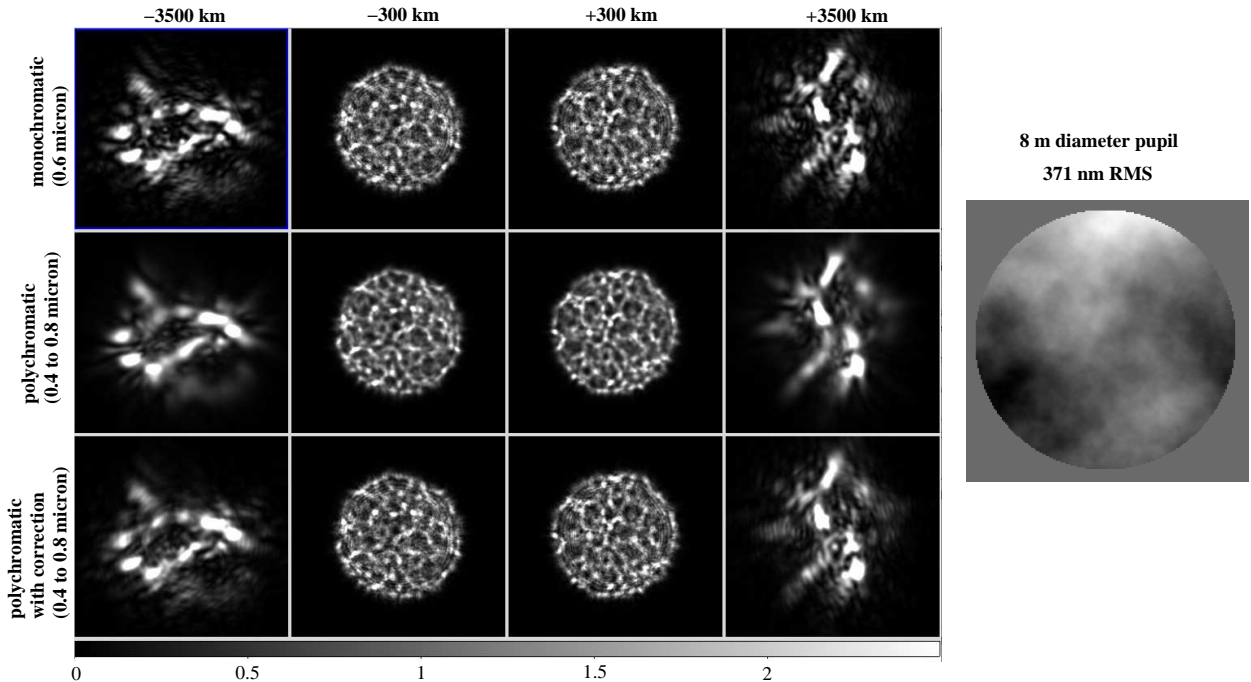}
\caption{\label{fig:defoc_chroma} Simulated frames obtained by the nlCWFS in monochromatic light (top), in polychromatic light with no chromatic compensation (center) and in polychromatic light with chromatic compensation (bottom). The wavefront error across the 8m diameter pupil used for this simulation is shown on the right.}
\end{figure*}

If non-linear effects are to be taken into account for tip-tilt measurement, the situation is radically different. First, larger defocus distances are allowed, which would improve the tip-tilt measurement sensitivity even if it were derived with the simple ``pupil photocenter'' diagnostic (which is linear). The real benefit however comes from the fact that the defocused pupil images are ``speckled'' due to edge (pupil edge, central obstruction, spiders) diffraction effects and high order aberrations. These highly contrasted speckles, visible in figure \ref{fig:defoc_chroma}, are $\lambda/D$ wide (which means they are physically smaller closer in to the pupil plane, and infinitely small - invisible -  at the pupil plane). Provided that the defocused pupil image is acquired with proper spatial sampling, a tip-tilt can therefore be measured as the displacement of a ``cloud'' of $\lambda/D$-wide speckles. High order aberrations must be known to some degree to be able to recover this highly sensitive tip-tilt measurement. Similarly, low order aberrations need to be known to measure high order aberrations from the defocused pupil images.

\subsection{Beam splitters}
\label{ssec:beamsplitters}
In conventional curvature systems, the required defocus is achieved sequentially with a resonant vibrating membrane (VM) in the system's focal plane. Because the high performance of the proposed wavefront sensing technique relies on resolving speckles in the defocused pupil images, care must be given to avoid smearing them, which would happen with a sine-wave motion of the membrane. If a VM were used, its motion should therefore be ``squared'', and the short transition period during which the defocus altitude is rapidly changing should not be used toward wavefront sensing (this time could conveniently be available for detector readout).

A better solution is to simultaneous image the four defocused pupil images on a single detector array (or on four smaller detector arrays) with a small number of beam splitters and mirrors. The desired beam size on the WFS detector array is on the order of 1mm (100 pixels accross the pupil for 10 $\mu$m pixels). In this scaled down copy of the telescope beam, a 1000 km vertical propagation distance is scaled down to approximately 1 cm.  In the non-linear curvature wavefront sensing scheme described in this paper, there is also no need to adapt the defocus altitudes as a function of atmoshperic conditions, so such a ``static'' implementation appears suitable.

\subsection{Chromatic compensation with the focal plane lenses}
\label{ssec:chromatlenses}
The primary purpose of lenses $L1+$, $L1-$, $L2+$ and $L2-$ in the WFS's focal planes is to achieve the desired altitude conjugation on the detector planes: if the lenses are removed, four identical pupil plane images would be acquired. In the nlCWF, these lenses are designed to be chromatic so that, in each of the 4 channels, the conjugation altitude is inversely proportional to wavelength. Figure \ref{fig:defoc_chroma} illustrates this chromatic compensation. Without compensation, speckles in the polychromatic images are slightly blurred. This is more apparent in with larger propagation distances, for which the finest-sized speckles disappear. With the chromatic compensation, polychromatic defocus pupil images are very close to monochromatic images, and the blurring of the smallest speckles is very minimal.
For very large propagation distances, defocused pupil images are close to a focal plane image, and the proposed chromatic compensation is equivalent compensating for the $\lambda$ scaling of the position of PSF diffraction features in the focal plane.

\subsection{Number of defocus planes}
\label{ssec:nbplanes}
In \citet{guyo08}, the dual stroke sensing scheme was justified by the need to mitigate some of the constraints brought by the requirement of preserving a linear relationship between WFS signals (curvature) and input wavefront phase. With a non-linear wavefront reconstruction scheme, it would therefore seem that the need for dual stroke is removed.

With the non-linear reconstruction algorithm and simulation parameters presented in this paper, the AO loop could not be closed using two symmetric detector planes. No attempt has been made to explore other configurations (for example, 2 non symmetric planes, or 3 planes) or stabilize the loop by modifying or constraining the reconstruction algorithm, so this does not exclude the possibility that nlCWFS may work with fewer than 4 planes.

While the reasons for the loop instability in a symmetric 2 planes nlCWFS have not been studied, it may be due to a fundamental property of diffractive propagation. The nlCWFS achieves its sensitivity by acquiring images at large propagation distances $z$ from the pupil plane. In such an image, there is a spatial frequency $f = \sqrt{2/(z \lambda)}$ for which $z$ is the Talbot propagation distance at which the intensity modulation vanishes. It may therefore be important to measure the pupil intensity at two different values of $z$ to ensure that every spatial frequency within the control range of the AO system produces an intensity signal in the WFS.

\subsection{Detector sampling}

The effect of detector sampling for nlCWFS has not been numerically quantified in this work, but physical considerations suggest it is driven by two requirements:
\begin{enumerate}
\item{The non-linear dual stroke curvature achieves its high sensitivity for low order aberrations by ``tracking'' $\lambda/d$ wide speckles in the defocused pupil images. The sampling therefore needs to be sufficiently fine to image these speckles with 2 pixels per $\lambda/D$ for maximum sensitivity. The linear number of pixels across the pupil should therefore be at least $N = 2 D^2/(\lambda z)$ where $z$ is the defocus distance and $D$ is the telescope diameter. With $D = 8m$, $\lambda = 0.8 \mu m$ and $z = 2000 km$, we find that the pupil should be 80 pixels across. Coarser sampling can be achieved by pushing the defocus altitudes higher, with little or no loss in performance since the reconstruction algorithm can accurately take into account non-linear effects.}
\item{As the defocus distance is increased, Fresnel diffraction ``blurrs'' the pupil image, which becomes larger than the geometric pupil. This effect is visible in fig. \ref{fig:defoc_chroma}. This is especially important for the highest spatial frequencies which the WFS must measure, since high spatial frequencies will diffract at larger distances away from the geometric pupil. The ``blurred'' pupil size becomes $D_{blurred} = D + 2 z \alpha$ where $\alpha$ is the greater of two values: $\alpha_1$, the angular ``control radius'' of the AO system at the sensing wavelength, or $\alpha_2$, half the angular size of the PSF at the wavefront sensing wavelength. In closed loop, $\alpha_2$ is usually much smaller than the natural seeing, and $\alpha_2$ is usually smaller than $\alpha_1$. With $D = 8 m$, $z = 2000 km$, $\lambda= 0.8 \mu m$, and $\alpha$ = 0.2\arcsec = 10 $\lambda/d$ (approximately 300 modes sensed), the detector needs to extend almost 2 m in every direction around the edge of the 8 m geometric pupil size.}
\end{enumerate}
When accounting for both effects, the linear number of pixels across the detector should be:
\begin{equation}
\label{equ:detecsampl}
N = \frac{2 D (D+2 z \alpha)}{\lambda z}.
\end{equation}
This equation clearly shows that smaller values of $z$ require more pixels. To minimize the number of pixels read, it is best to chose $z$ equal to or greater than $z_0 = D/(2 \alpha)$, which is the propagation distance for which the ``blurred'' pupil size is twice the geometric pupil size. For $D = 8m$ and $\alpha$ = 0.2\arcsec, $z_0 = 4000 km$. For $z=z_0$, $N = N_0 = 4 D \alpha / \lambda$, and for $z >> z_0$, $N = N_{\infty} = N_0/2$. 

With $z >> z_0$, the total number of pixels required per conjugation plane is $N_{\infty}^2 \pi/4 = \pi D^2 \alpha^2 / \lambda^2$, which is equal to the number of $\lambda/D$-wide speckles within the control radius $\alpha$ of the AO system. Since each of these focal plane speckes corresponds to two degrees of freedom (amplitude and phase), a pair of defocused pupil images contains as many pixels as there are modes sensed by the nlCWFS. With two such pairs of defocused images read, and if $z > z_0$, the nlCWFS therefore requires at least 2 pixels to be read per WFS mode. Compared to a Shack-Hartmann WFS with 2x2 pixels cells or a pyramid WFS designed to sense the same total number of modes, the non-linear dual stroke curvature WFS is therefore expected to require as many or slightly more pixels to be read.

Equation \ref{equ:detecsampl} has not been verified by numerical simulations, and the non-linear iterative nature of the reconstruction may produce a more complicated picture than the one presented in this section. Both the required detector sampling and the required number of conjugation planes are important drivers for system cost and complexity, and will need to be explored in the future by comparing closed loop numerical simulations for several combinations of defocus distances, pixel scale for each plane, and detector size for each plane. Reconstruction algorithms should also be chosen to mitigate coarse detector sampling or limited number of conjugation planes if possible.

\section{Wavefront reconstruction algorithm}

\label{sec:algo}

In this section, an iterative algorithm which successfully reconstructs the wavefront from the four defocused pupil images is identified. While more efficient algorithms may exist for this problem, this algorithm is then used in the rest of this paper to estimate the nlCWFS performance.

\begin{figure*}
\includegraphics[scale=0.37]{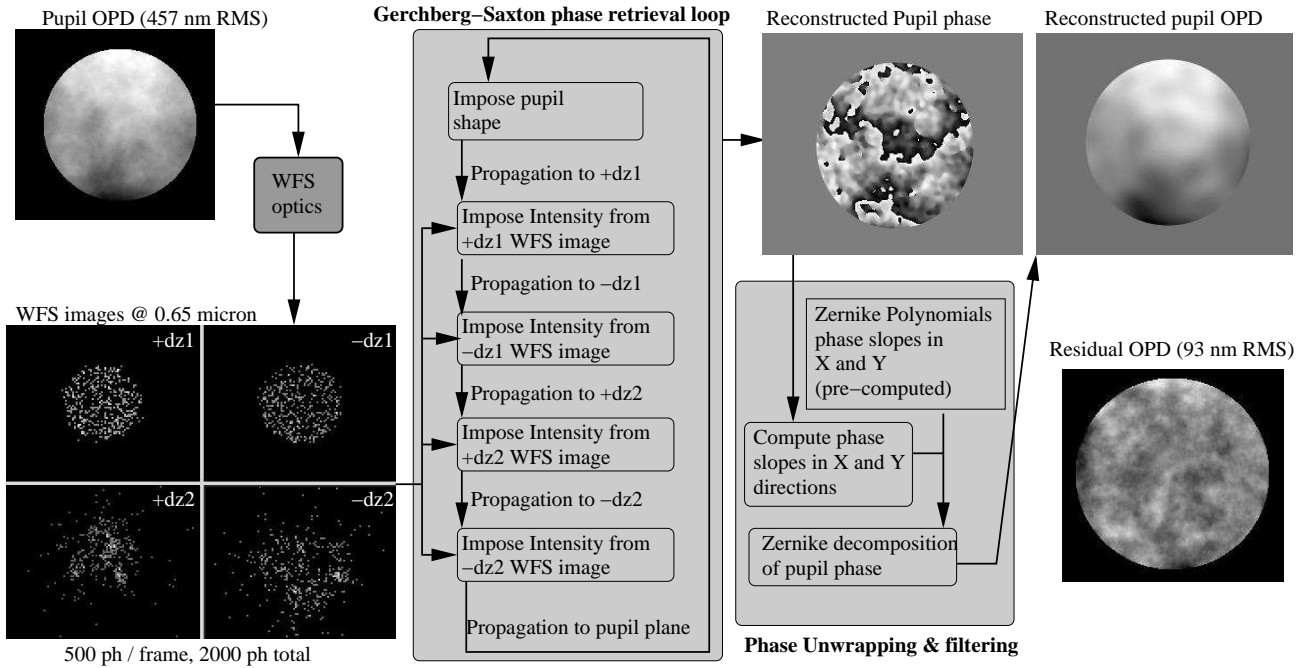}
\caption{\label{fig:wfsalgo} Proposed wavefront reconstruction algorithm for non-linear curvature wavefront sensing.}
\end{figure*}

\subsection{Phase retrieval}
\label{ssec:phaseretr}
The wavefront may be reconstructed from several defocused pupil images through a phase diversity algorithm \citep{gons79}. A Gershberg-Saxon \citep{gers72} algorithm is chosen in this work for simplicity and flexibility, but many other options and variations on this algorithm are available \citep{fien82}, as much work has been devoted to phase retrieval algorithms in the last few decades. Similar phase retrieval algorithms have been successfully used to measure optical aberrations in the Hubble Space Telescope \citep{fien93,rodd93,kris95}. The algorithm used for this work is illustrated in figure \ref{fig:wfsalgo}.


Using this algorithm, nearly optimal closed loop performance has been obtained in simulations with as few as 10 GS loops by (1) Òfast forwardingÓ the convergence thanks to the fact that in consecutive GS iterations, the solution moves in the same ÒdirectionÓ, and (2) running a few lower resolution GS iterations on binned data to rapidly approach the solution prior the full size slower GS iterations. Essentially all of the computation time is taken by the full-size FFTs in the propagations, while the phase unwrapping described in the next section is very quick, as it is both linear and non-iterative (see \S\ref{ssec:unwrap}).
Execution speed is a steep function of sampling of the defocused pupil images. Eight complex 64x64 pixels 2D FFTs are required per GS iteration (assuming a 45 pixel diameter pupil). On a single modern CPU, each FFT takes 0.088 ms (number adopted from http://www.fftw.org/speed/ for a 3 GHz Intel Core Duo CPU), corresponding to 7ms total FFT time. On a single CPU, the iterative nlCWFS discussed in this section can therefore run at $\approx$100 Hz with current hardware.

Other approaches to this non-linear problem exist, but have not been tested in this work. \cite{vand02c} have suggested a geometry-based algorithm to derive wavefront aberrations from defocused image. The reconstruction can also be linearized in a $\approx 1 rad$ domain around any chosen wavefront, and it should therefore be possible to effectively combine a coarse non-linear algorithm with a fine resolution linear algorithm.


\subsection{Phase unwrapping and filtering}
\label{ssec:unwrap}
The output of the GS iterative algorithm is a phase map of the pupil, which then needs to be unwrapped and filtered before commands are sent to the DM. Unwrapping may not be necessary if the closed loop performance is such that the peak-to-valley phase is less than $2 \pi$. Figure \ref{fig:wfsalgo} shows how unwrapping and filtering (in this example, projection on a set of Zernike polynomial) can be done together in one step. The phase slope is insensitive to phase wrapping effects, as it is here computed as difference between phase of adjacent pixels, and any value above $\pi$ or below $-\pi$ is brought back in the $-\pi$ to $\pi$ interval by adding or subtraction $2\pi$. For each of the two axis (x and y) the resulting phase slope is decomposed as a linear sum of precomputed phase slopes (one per Zernike polynomials). The coefficients of this decomposition are the Zernike coefficients of the wavefront, which can therefore simultaneously be unwrapped and filtered by being reconstructed as a sum of Zernike polynomials. If the output of the GS algorithm is sufficiently good, the difference between the wrapped pupil phase and the filtered/unwrapped pupil phase (this difference is small enough to be free of phase wrapping effects) may be added to the final unwrapped phase map to get back high spatial frequencies removed by the filtering step. If however the WFS frames are very noisy, some filtering may also be included within the GS loops to force the GS algorithm to produce a smooth phase map, which will mitigate problems associated with phase unwrapping errors on noisy wavefront data.
The final step of the wavefront estimation is to appropriately weigh the output of our wavefront reconstruction algorithm with prior knowledge of the wavefront and knowledge of the statistical properties of the wavefront \citep{gend94,irwa98} and its temporal evolution. This optimization problem has not been explored in this paper, and is not specific to the nlCWFS wavefront sensor (although the exact ``tuning'' of this optimization is wavefront sensor-dependent).

\begin{figure*}[htb]
\includegraphics[scale=0.67]{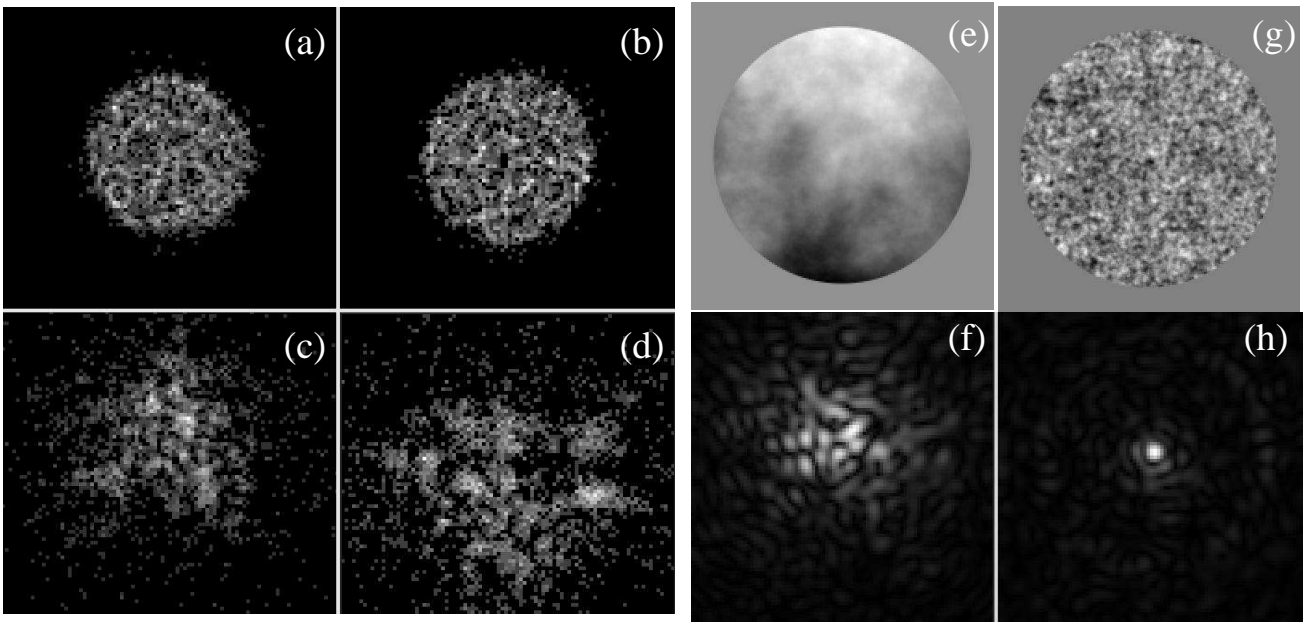}
\caption{\label{fig:609nm} Wavefront reconstruction using the algorithm shown in fig. \ref{fig:wfsalgo}. Four noisy defocused pupil images (images (a), (b), (c) and (d)) are acquired to transform the pupil phase aberrations (e) into intensity signals. The input pupil phase is 609 nm RMS, yielding the PSF (f) before correction. After correction, the residual pupil phase aberration (g) is 34.4 nm RMS, allowing high Strehl ratio imaging (h). All images in this figure were obtained at 0.65 $\mu$m. The total number of photons available for wavefront sensing in 2e4.}
\end{figure*}

\section{Performance: Numerical Simulations}
\label{sec:perf}
In this section, numerical simulations are performed to verify and evaluate the nlCWFS's performance in a closed loop adaptive optics system. In \S\ref{ssec:brightstars}, reconstruction of a single wavefront is tested using the algorithm shown in Fig. \ref{fig:wfsalgo}. Closed loop simulations of nlCWFS and SH based AO systems are presented in \S\ref{ssec:clssh} to illustrate the sensitivity gain offered by nlCWFS. Chromaticity effects are investigated in \S\ref{ssec:chrom} with and without the optical chromaticity compensation proposed in this work. In \S\ref{ssec:faint}, the flux limit at which nlCWFS starts to loose the ability to see individual speckles is probed through numerical simulations. Finally, in \S\ref{ssec:sparse}, closed loop simulations are shown for sparse apertures.

\subsection{Reconstruction of a single aberrated wavefront}
\label{ssec:brightstars}
Figure \ref{fig:609nm} illustrates the performance of non-linear CWFS. For this simple simulation, the initial wavefront aberrations were 609nm RMS across the pupil. This level of aberration is, for a 8m telescope, intermediate between what is expected to be encountered in open loop (typically 2000nm RMS) and in closed-loop (around 100nm RMS for a bright guide star). The number of photons available for wavefront sensing was chosen to be 2e4, which corresponds to approximately 0.5 ms integration time on a $m_V=10$ source in a $0.5 \mu m$ wide bandwidth and a 20\% system efficiency (product of optics throughput and detector quantum efficiency). This example is therefore very representative of situations commonly encountered in ground-based Adaptive Optics when observing moderately bright targets.

Results shown in figure \ref{fig:609nm} demonstrate the nlCWFS's good sensitivity: with only 2e4 photons (photon noise is very pronounced in the WFS frames), the wavefront is reconstructed to a 34nm RMS accuracy. The PSF obtained is therefore diffraction limited in the visible. In the next part of this paper, simulations are performed in conditions closer to a real AO system: in \S\ref{ssec:chrom} the effect of chromaticity is evaluated, and in \S\ref{ssec:clssh} and \S\ref{ssec:faint} detailed closed-loop simulations are used to verify performance.

\subsection{Closed-loop simulation: performance and direct comparison with SHWFS}
\label{ssec:clssh}
\begin{figure*}[htb]
\includegraphics[scale=0.44]{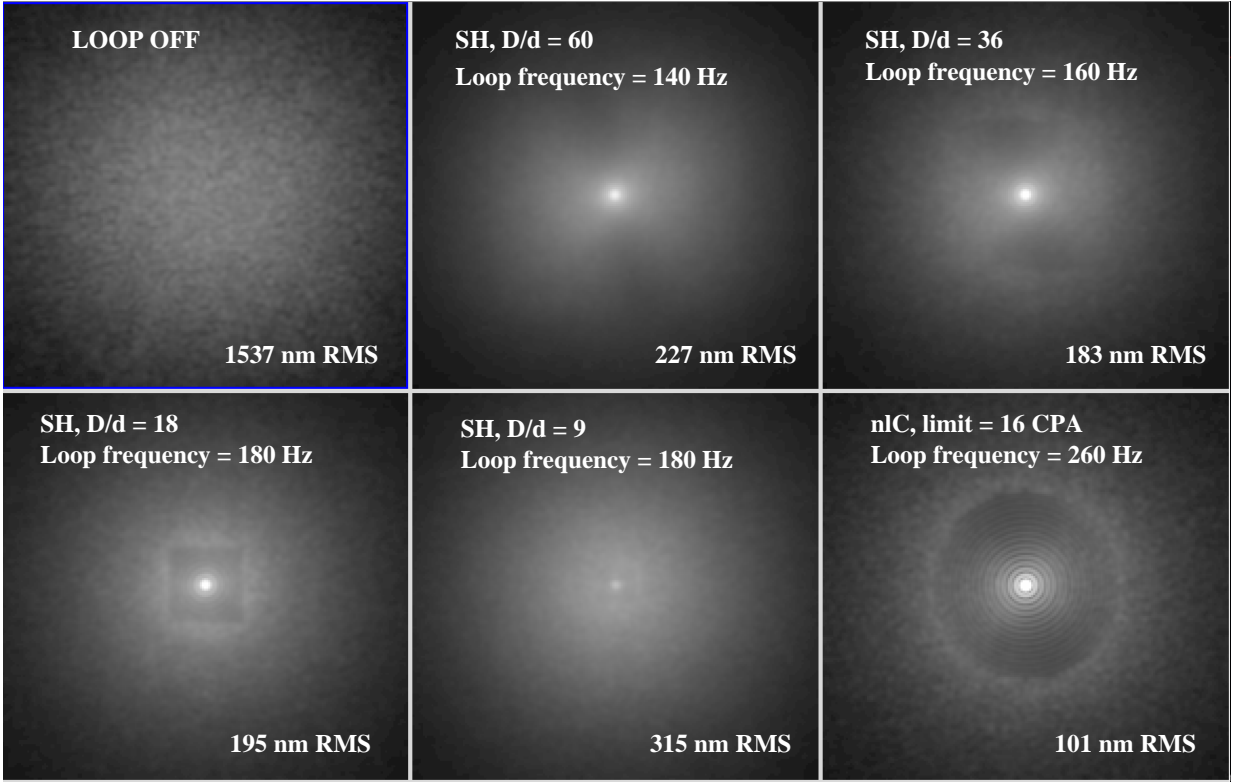}
\caption{\label{fig:SHvsnlC16} Simulated 10 second exposures closed loop PSFs delivered by AO systems with SHWFS and nlCWFS.}
\end{figure*}

Figure \ref{fig:SHvsnlC16} compares closed loop AO performance obtained with SHWFS and nlCWFS for a total flux of 3x$10^6$ ph/s on a 8-m telescope (magnitude ~13 with 20\% efficiency and a 0.5 $\mu$m wide bandpass). Parameters adopted for this simulation are given in Table \ref{tab:AOsimparam}. Both the WFS and imaging wavelength are 0.85 $\mu$m. The WFS detector array is considered perfect (no readout noise, no dark current). For the SH systems, each subaperture PSF is imaged on 64x64 pixels, and the DM is considered ideal, correcting all spatial frequencies measured. Four SHWFS configurations were simulated with D/d varying from 60 to 9 (D = telescope diameter, d = subaperture size). For the nlCWFS simulation, the detector acquires simultaneously the four defocused pupil images with 90 pixels across the pupil in each image. The DM is assumed to perfectly correct all modes up to 16 cycles per aperture (CPA), and provides no correction for higher spatial frequencies.

\begin{deluxetable}{lccc}
\tabletypesize{\small}
\tablecaption{\label{tab:AOsimparam} Default parameters used for simulations}
\tablewidth{0pt}
\startdata
\hline
& nlCWFS & SHWFS\\
\hline
Telescope diameter & \multicolumn{2}{c}{8m} \\
Seeing & \multicolumn{2}{c}{0.6\arcsec FWHM at 0.5 $\mu$m} \\
Wind speed & \multicolumn{2}{c}{10 m.s$^{-1}$}\\
WFS wavelength & \multicolumn{2}{c}{0.85 $\mu$m} \\
WFS number of subapertures & not applicable & from 9 to 60 subapertures across pupil \\
WFS defocus distances & -3500 km, -2500 km, 2500 km, 3500 km & not applicable\\
WFS detector sampling & 90 pixels across pupil & 64x64 pix. per subaper. (pix. size $\ll \lambda/d$) \\
WFS detector readout noise & \multicolumn{2}{c}{0} \\
AO loop gain &  \multicolumn{2}{c}{1.0} \\
AO loop modal filtering & \multicolumn{2}{c}{none (gain = 1 for all modes within control range)} \\
Spatial frequency control range & up to 16 cycles per aperture & up to CPA $= D/3d $\\
Edge subapertures & not applicable & active if $>$ 70 \% illuminated\\ 
\enddata
\end{deluxetable}

The simulations code was written in C language, and used the fftw library \citep{fftw} for fast Fourier transforms. In each case, AO loop frequency was optimized to minimize residual wavefront aberration (RMS phase error over the pupil): this optimization was done by running the loop at different frequencies in 20Hz increments and selecting the optimal solution. Figure \ref{fig:SHvsnlC16} illustrates the effect of noise propagation in a SHWFS: with a large number of subapertures (D/d = 60), sensing accuracy for low order aberrations is poor, and the overall wavefront RMS error is large even though a high number of modes is corrected. With poor wavefront sensing sensitivity, the optimal loop frequency is low (140 Hz for D/d=60) to balance time lag and photon noise. With fewer larger subapertures (D/d = 36), correction is better for low spatial frequencies, and residual wavefront error drops even though high spatial frequencies are not corrected, as can be seen from the light just outside the darker square indicating the control region of the AO system. Further reducing the number of subaperture however fails to significantly improve correction of low spatial frequencies as the SHWFS sensitivity is now dominated by seeing size (the subapertures are larger than $r_0$). For a given guide star brightness, there is an optimal number (higher number for brighter stars) of subapertures in a SHWFS for which sensitivity loss on low order aberrations is balanced against number of modes measured.

Thanks the nlCWFS's high sensitivity across the full range of spatial frequencies measured, the PSF halo is much fainter than could be achieved with any of the SHWFS configurations tested. This improved sensitivity also allows the nlCWFS-based AO system to run faster (260 Hz). Increasing the number of wavefront modes measured does not lower performance for low order aberrations in a nlCWFS.

\subsection{Chromaticity}
\label{ssec:chrom}

As described in \S\ref{ssec:chromatlenses}, the proposed design nlCWFS uses chromatic lenses to improve the wavefront sensor's ability to work in broadband light. This scheme does compensate for chromaticity induced by diffraction propagation, but will blur the defocused pupil images in the geometrical optics regime where aberrations are large. For example, a large amount of tip/tilt $\alpha$ will shift defocused pupil images by $\alpha dz$ independently of wavelength. The chromatic lenses proposed in \S\ref{ssec:chromatlenses} will blur the images since $dz$ is now a function of $\lambda$. It is therefore important to verify that the use of chromatic lenses does not prevent the loop from initially closing when aberrations are large, and that the correction remains stable in time.

\begin{figure}[htb]
\includegraphics[scale=0.37]{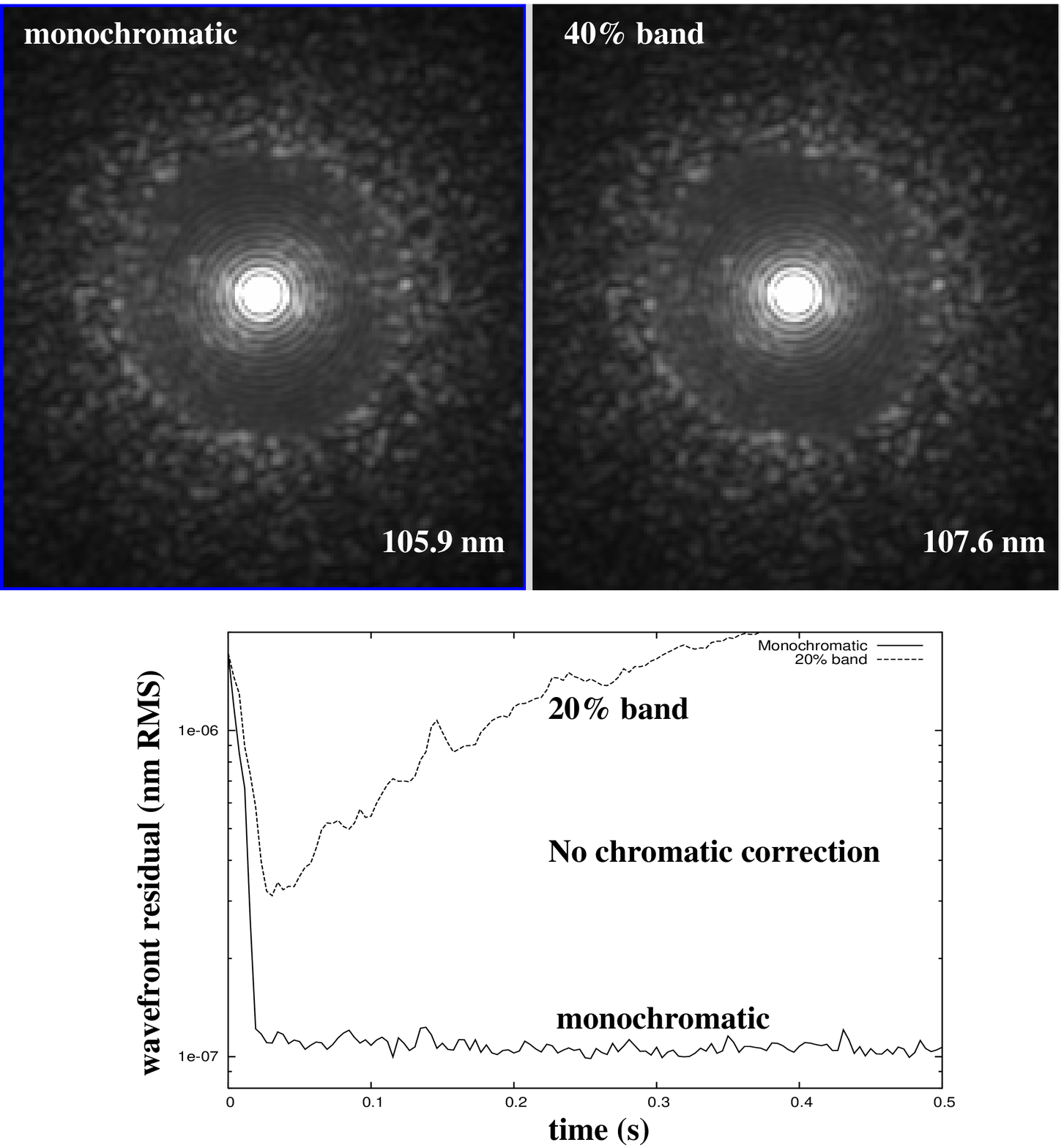}
\caption{\label{fig:chromat}Top:  Closed loop 1 second exposure PSFs at 0.85 $\mu$m with a nlCWFS working in monochromatic and broadband (d$\lambda$/$\lambda$ = 40\%) light. In both simulations, the total number of photons in the nlCWFS is 3e6 ph/s. Bottom: Without chromatic correction within the nlCWFS (replacing the chromatic lenses by achromatic lenses in Figure \ref{fig:nlcurvaoprinc}), the loop would be unstable with d$\lambda$/$\lambda$=0.2.}
\end{figure}

Simulation results shown in Figure \ref{fig:chromat} confirm that the chromatic lenses scheme allows the nlCWFS to both close the loop and maintain a good correction in broadband light with little loss of performance compared to the monochromatic case. Simulations shown in Figure \ref{fig:chromat} use the same parameters as used in the nlCWFS result shown in Figure \ref{fig:SHvsnlC16}, except for dz1, which was changed from 2500 km to 300 km. Reducing dz1 does help with loop stability in polychromatic light, at very little cost in performance. Without the chromatic lenses, the loop would not have been stable in a 20\% wide band (Figure \ref{fig:chromat}, bottom).

These results are very encouraging, as they were obtained by using a wavefront reconstruction algorithm which assumes the nlCWFS is working in monochromatic light. It therefore appears that polychromatic operation of a nlCWFS requires no additional computing power. 

Broadband use of the nlCWFS also requires an atmospheric dispersion compensator (ADC) ahead of the nlCWFS. For nlCWFS to operate at full efficiency, chromaticity introduced by atmospheric dispersion should be accurately compensated to keep the lateral PSF shift below the diffraction limit of the telescope across the wavefront sensing spectral coverage: even a small angular extent of the source used for wavefront sensing is highly detrimental.

\subsection{Coherence loss and optimal Wavefront Sensing wavelength}
\label{ssec:faint}

\begin{figure}[htb]
\includegraphics[scale=0.52]{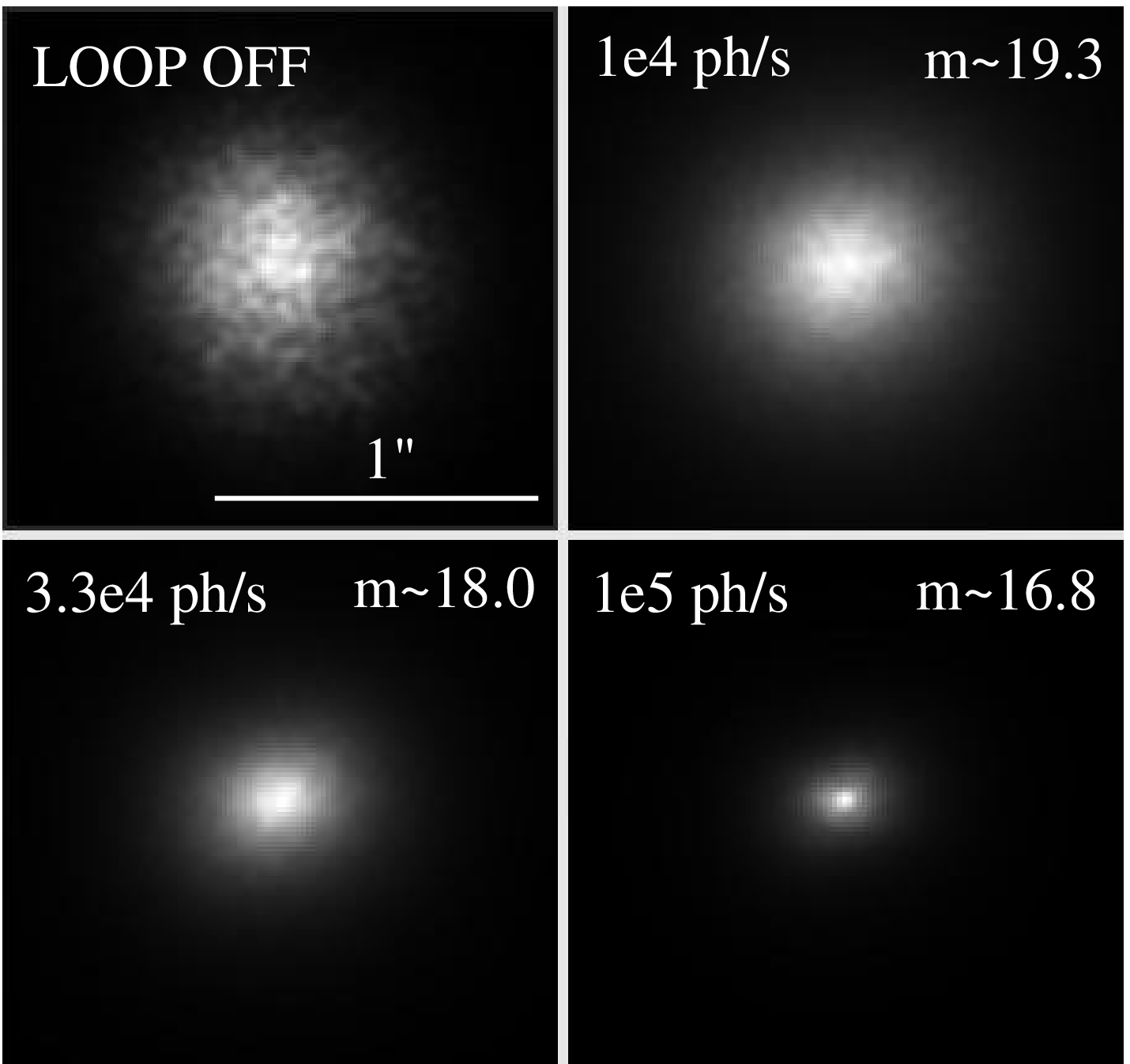}
\caption{\label{fig:FaintPSFs} Simulated long exposure 1.6 $\mu$m PSFs obtained with a non-linear Dual stroke Curvature-based AO system. The sensing wavelength is 0.85 $\mu$m for this simulation.}
\end{figure}

\begin{figure*}[htb]
\includegraphics[scale=0.6]{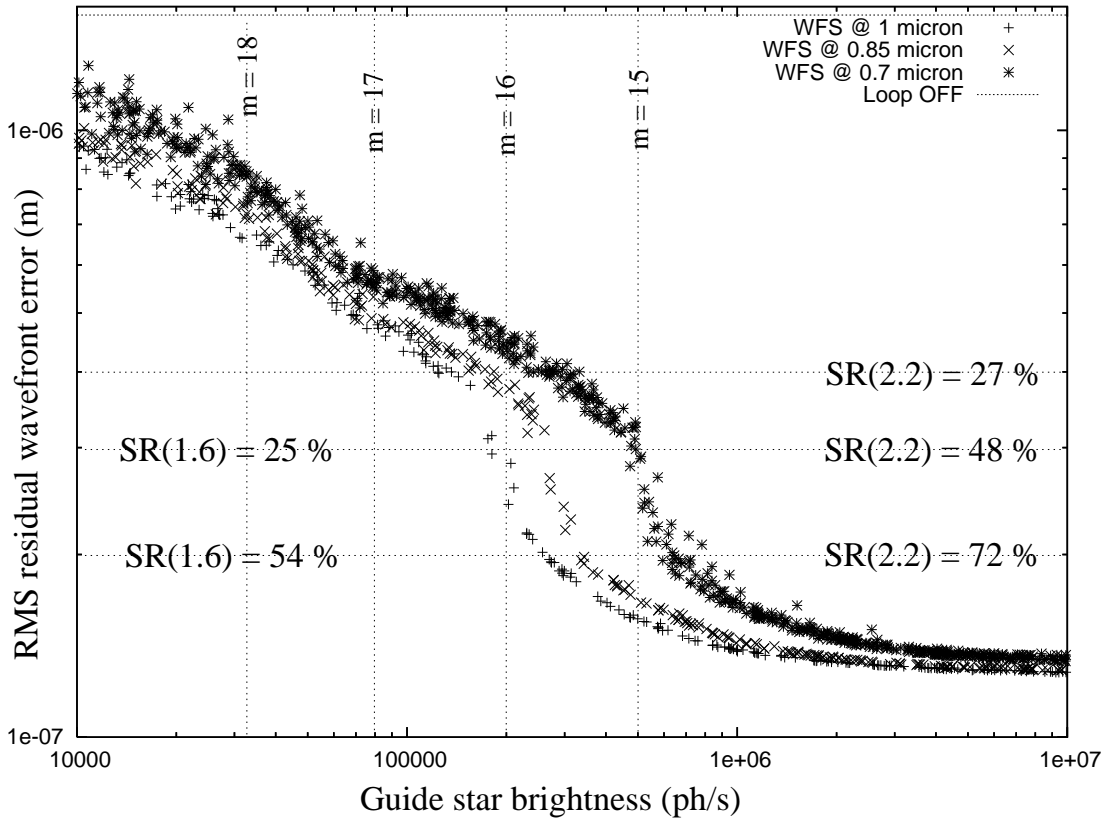}
\caption{\label{fig:RMSperfplot} Simulated performance of a low order nlCWFS-based system (first 200 Zernike modes corrected) as a function of sensing wavelength (0.7, 0.85 and 1.0 $\mu$m) and guide star brightness. The stellar magnitudes given in this figure assume a 20\% efficiency in a 0.5$\mu$m wide band. Each point in this figure is a 1-second average of the RMS residual wavefront error. See text for details. }
\end{figure*}

Closed loop simulations using the nlCWFS show a rapid drop in performance at about $m \approx$15 (Figure \ref{fig:RMSperfplot}). This limit is due to a loss of coherence in the pupil plane: the number of photons becomes too low to ÒtrackÓ the diffraction-limited speckles which are essential to the nlCWFSÕs sensitivity. Beyond this limit, the nlCWFS operates in a similar regime as conventional linear WFSs, and delivers partially corrected PSFs (Figure \ref{fig:FaintPSFs}). 

The results obtained in figure \ref{fig:RMSperfplot} also show that for faint guide stars (approximately $14<m_V<17$ in the figure), it is preferable to choose a longer wavefront sensing wavelength, thanks to the recovery of coherence in the pupil plane. For bright stars, a shorter sensing wavelength is better, since the same OPD will correspond to a larger phase shift, which can be detected more easily. Figure \ref{fig:RMSperfplot} does not show this last point because the RMS residual wavefront error for bright stars is dominated by the number of modes corrected in this "low-order" AO system.

\subsection{Sparse apertures, spiders and low order aberrations}
\label{ssec:sparse}

\begin{figure*}[tbh]
\includegraphics[scale=0.435]{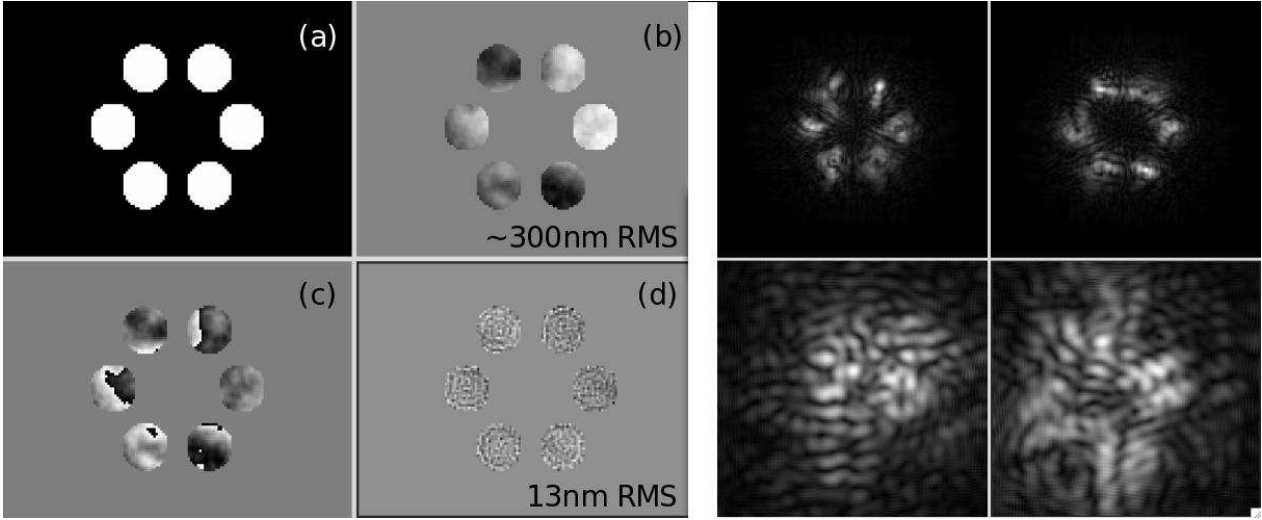}
\caption{\label{fig:sparsepup} Wavefront reconstruction with dual stroke non-linear CWFS on a sparse pupil. The pupil amplitude (a) and phase (b) yield the four defocused pupil images shown on the right. The recovered pupil phase (c) and the residual phase error (d) demonstrate that dual stroke non-linear CWFS can simultaneously measure OPD within and across segments. This polychromatic simulation was performed with 2e8 photons in a $d\lambda/\lambda=0.5$ wide band centered at 0.65 $\mu m$.}
\end{figure*}

For both existing and future telescopes, the pupil is usually not a single continuous disk as assumed so far in this work. Narrow spider vanes have a very small effect on most wavefront sensing schemes, but thicker spider vanes can be a more serious problem, especially if a Shack-Hartmann sensor is used with subaperture size comparable to or smaller than the spider thickness. This situation will likely occur on the next generation of large telescopes. This problem is even more serious for a large telescope composed a few large segments, with large gaps between the segments, of which the Giant Magellan Telescope is the perfect example. Incorporating statistical knowledge of the phase error can help, but will not work for large gaps or vibration-induced piston between segments. A separate scheme to measure differential piston between segments is often required to solve this problem.

To illustrate this problem, I now consider a telescope with 20cm thick spider vanes, and a Shack-Hartmann wavefront sensor. To improve the AO system performance for bright stars, a large number of small subapertures is preferred. With ~1m subapertures, the spiders would not be a serious problem, but with ~20cm subapertures, the subapertures along the spiders are either fully dark or partially illuminated. No subaperture sees light simultaneously from both sides of the spider and the WFS is therefore blind to an OPD between the two sides of the spider. Even if the subapertures are twice the spider thickness, the amount of light which is mixed between the two sides of the gap is relatively small, and OPD measurement between the two sides will be noisy. The ability to measure wavefront across a gap is a function of how much light is mixed within the WFS between the two sides of the gap. Shack-Hartmann sensors, especially with small subapertures, are therefore poorly suited for this task.

As illustrated in figure \ref{fig:sparsepup}, with a nlCWFS the images acquired are sufficiently far from the pupil plane to introduce significant mixing between pupil ``segments'', therefore enabling high sensitivity measurement of OPD between segments. 
A fundamental problem however persists if the gaps are wide: no monochromatic wavefront sensing scheme can detect OPDs which are multiples of $\lambda$. The solution to this problem (which has not yet been explored) is to intelligently use chromatic effects: a trial and error iterative algorithm using the focal plane science image (which, presumably is at a different wavelength) may be used to ``lock'' the inter-gap OPD within $\lambda$ - once this is achieved, the closed loop nlCWFS-based  AO system will accurately maintain the correct inter-gap OPD. Alternatively, the broadband defocused pupil images acquired within the WFS should be sufficient to solve for this problem: the ``interference'' fringes between the segments are highly contrasted if the inter-gap OPD is close to zero, but much less contrasted if it is close to $\lambda$. Using this as a diagnostic could reveal the absolute value of the intergap OPD, not its sign: some iterative trial-and-error algorithm seems necessary to initially lock the AO control loop in the correct intergap OPD.

The fundamental gain (over a Shack-Hartmann sensor) offered by non-linear CWFS for sensing inter-gap OPD measurement is merely a demonstration of its higher sensitivity for low-order aberrations sensing. An inter-gap OPD is mostly a low-order aberration. In both cases (inter-gap OPD and low order aberration), the key to high sensitivity lies in being able to mix light from relatively distant parts of the pupil.

\section{Discussion}

\label{sec:disc}

\subsection{Sensitivity comparison with Other Adaptive Optics Wavefront Sensing Techniques}
\label{ssec:senscomp}
\begin{figure*}
\includegraphics[scale=0.33]{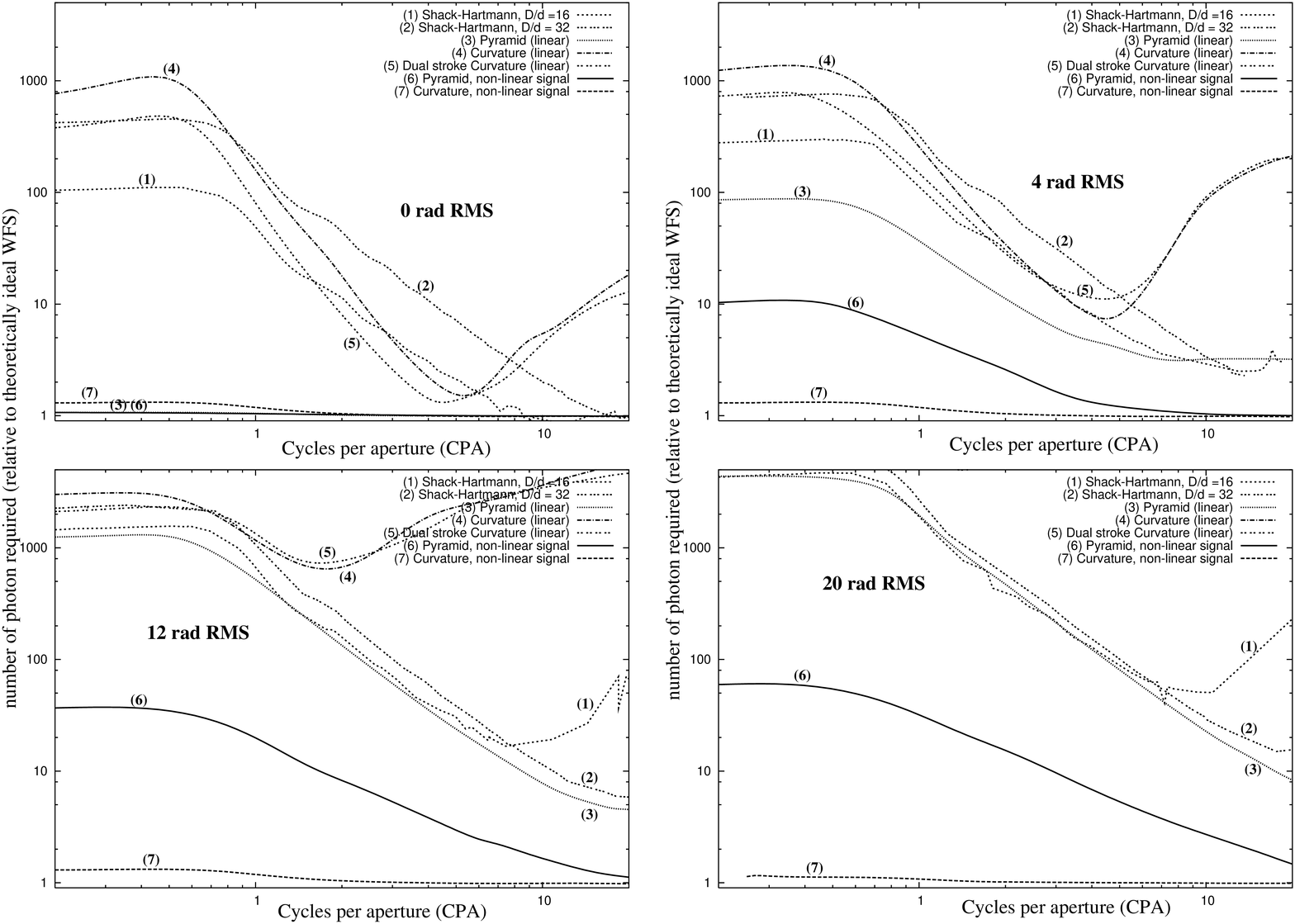}
\caption{\label{fig:opticalsens} Optical sensitivity of several wavefront sensor options. The number of photon (y axis, relative) required to achieve a fixed wavefront sensing SNR is shown as a function of spatial frequency (x-axis). Sensitivities are shown for wavefront quality ranging from perfectly flat (top left) to 20 rad RMS (bottom right).}
\end{figure*}

Figure \ref{fig:opticalsens} shows for different wavefront sensing schemes how strong an optical signal is produced when small changes are introduced in the wavefront of a 8 m diameter telescope. The WFS is operating at 0.85 $\mu$m, and all phases are measured at this wavelength. The "strength" of the signal is measured as signal to noise (SNR) for a fixed number of available photons, or, equivalently, how many photons are required to reach a fixed SNR. Although only photon noise is considered, the highest sensitivity WFSs will also be more robust against detector readout noise.
The WFS schemes compared in Figure \ref{fig:opticalsens} are:
\begin{itemize}
\item{Two SHWFSs with different number of subapertures ($D/d = 16$ and $D/d = 32)$, both using linear wavefront reconstruction.}
\item{A curvature WFS with a $dz=1000 km$ propagation distance on either side of the pupil plane and a linear wavefront reconstruction}
\item{A linear dual-stroke curvature WFS \citep{guyo08} with $dz1 = 800 km$ and $dz2 = 2000 km$}
\item{A linear pyramid wavefront sensor where the pyramid position is fixed (Fixed Pyramid WFS, or FPYRWFS).}
\item{A FPYRWFS where non-linear signal is taken into account for wavefront reconstruction sensitivity.}
\item{A curvature WFS where non-linear signal is taken into account, with two defocus distances $dz1 = 2500 km$ and $dz2 = 3500 km$ (4 pupil images total).}
\end{itemize}
The y-axis shows how many photons are needed to achieve a fixed wavefront sensing signal-to-noise ratio (SNR) compared to the optimal sensitivity wavefront sensor as defined in \cite{guyo05}. For example, the top left panel (0 rad RMS case) shows that at 1 cycle per aperture (CPA), both the SHWFS with $D/d = 32$ and the curvature WFS require ~200 times more photons than an ideal WFS to produce an intensity signal with the same SNR. This flux factor is given as a function of spatial frequency (x axis), measured here in CPA.
Figure \ref{fig:opticalsens} was generated by (1) Computing the intensity response in the WFS, (2) Computing how this intensity response changes when a small aberration is added to the wavefront, and (3) Comparing this change, pixel by pixel, to the photon noise and combining the resulting SNRs into a single overall SNR. This computation was done not only for a flat incoming wavefront, but also for aberrated wavefronts, and therefore quantifies optical sensitivity beyond the small perturbation linear regime considered in \citet{guyo05}. A distinction is also made between ÒlinearÓ schemes (SH, conventional curvature, conventional pyramid), where the intensity signals from a series of uncorrelated wavefronts were added, and Ònon-linearÓ schemes, where a single aberrated wavefront is used. This computation step is necessary to average out non-linear signals in linear WFSs: for example, a SHWFS cannot use the shape and speckles of individual spots to unambiguously measure high order aberrations within each subaperture; and the centroids measurement error is driven by the overall size of the spots instead of the size of speckles within these spots

Figure \ref{fig:opticalsens} shows why a Curvature WFS in which non-linear signal is used appears to be the most promising approach. Other existing WFS designs cannot as easily be modified to offer similar benefits:
\begin{itemize}
\item{The SHWFS suffers from poor sensitivity at low spatial frequencies. This poor sensitivity is fundamentally due to the splitting of the pupil into subapertures which prevents interferences between distant points in the pupil.}
\item{The linear fixed pyramid WFS is very sensitive at all spatial frequencies in the small aberration regime, but shows poor sensitivity at low spatial frequencies when the wavefront quality is less than ideal. This originates from both the linear constraint and an optical range limitation due to the fact that the inner parts of the PSF can spend a significant fraction of the time completely in a single quadrant of the pyramid - in this configuration, the WFS does not produce a strong optical signal for low order aberrations. If the non-linear part of the signal is used, sensitivity is improved, but is still, at low spatial frequencies, about an order of magnitude poorer than it would be for a flat wavefront.}
\item{The linear curvature WFS lacks sensitivity at low spatial frequencies. The dual-stroke scheme offers a modest (factor 2) sensitivity improvement at low spatial frequencies when wavefront errors are moderate. As shown in \citet{guyo08}, in the defocused pupil images, mixing of light between distant parts of the pupil occurs, but it requires large defocus distances to be easily seen and it is highly non-linear. If this non-linear signal is taken into account to derive the WFS sensitivity, the dual-stroke CWFS is highly sensitive at all spatial frequencies when the wavefront is highly distorted. Curiously, simulations show that sensitivity is poorer when the wavefront quality is good: in this case, the speckles which carry the non-linear information in the defocused pupil images are weak or absent. In this regime, a static known aberration should be added in the WFS to maintain full sensitivity. A 2 $\lambda$ RMS static aberration with a Kolmogorov power spectral density has been included in Figure \ref{fig:opticalsens} in the 0 rad and 4 rad cases.}
\end{itemize}

This analysis therefore shows that the non-linear curvature wavefront sensor (nlCWFS) is, among all the approaches listed, the only one to offers near-ideal optical sensitivity at all spatial frequencies and all levels of wavefront aberrations within the 0 to 20 rad RMS range considered. The simulations shown in this section only considered the effect of a single wavefront mode at a time, and the results obtained do not automatically mean that the nlCWFS (or non-linear pyramid) signals can unambiguously be used to accurately measure the wavefront. For example, different wavefront aberrations could produce identical nlCWFS signals. The existence of a wavefront reconstruction algorithm which can recover a complex wavefront map was shown in \S\ref{sec:algo}.


\subsection{Comparison with phase diversity}
In Phase Diversity (PD), images near the focal plane are acquired to measure wavefront aberrations. Similarly to nlCWFS, a non-linear reconstruction algorithm is used to estimate pupil phase from several images acquired in different planes. The position of these planes is usually given in pupil defocus phase for PD while it is given in propagation distance for nlCWFS. Physically, an image of the pupil plane acquired after a free space propagation is equivalent to a defocused focal plane image. The two quantities are related by the following equation:
\begin{equation}
d\phi = \frac{\pi d^2}{4 \lambda l}
\end{equation}
where $d\phi$ is the pupil defocus term in radian peak-to-valley, $d$ is the telescope diameter and $l$ is the propagation distance.

In conventional linear curvature wavefront sensing, the propagation distance $l$ is about 1000 km or less, equivalent to at least 10 waves (peak to valley) of defocus for PD. PD operates at smaller defocus values, typically 1 to 5 waves, and it has been shown that operating PD with larger defocus values reduces its sensitivity \citep{fien98}.

In the nlCWFS example adopted for Figure \ref{fig:opticalsens}, $l$ is 3500 km, yielding $d\phi$ = 2.3 waves (peak to valley). nlCWFS therefore acquires images which are physically very similar to PD images, and erases much of the fundamental difference previously identified between PD and curvature wavefront sensing \citep{fien98}. The results presented in this paper, and especially Figure \ref{fig:opticalsens}, are consistent with the sensitivity comparison between PD and curvature previoulsy published in \cite{fien98}. One key finding of their study is that curvature WFS's sensitivity is fundamentally limited by the large defocus value (= small propagation distance in the pupil plane) at which it operates (as shown in Figure 5 in \cite{fien98}). This limitation is in fact due to the linearity constraint in conventional curvature, but is lifted in the nlCWFS scheme which allows non-linear signals to be used at large propagation distances. Figure 5 of \cite{fien98} shows that optimal sensitivity is reached when the defocus is below $\approx$ 5 waves for a wavefront aberration including the 52 lowest order Zernike polynomials. With higher order aberrations included and a well sampled detector, the same figure would show that the defocus can be increased further (the propagation distance can be reduced) because the acquired images would still contain the $\lambda/D$-wide speckles necessary for high sensitivity wavefront sensing.

\subsection{nlCWFS performance for Extreme-AO on bright targets}
\label{sec:exao}

\subsubsection{nlCWFS linearity in small aberration regime}
When wavefront variations are small (less than 1 rad), the defocused pupil images acquired by the nlCWFS become less contrasted and they are dominated by diffraction ringing from the pupil's edges: a known static aberration should then be added to the beam in the WFS path to maintain full sensitivity (see \S\ref{ssec:senscomp}). If the wavefront aberrations around this static wavefront offset are small ($<$1 rad), the nlCWFS is also linear.

In this low aberration regime, the nlCWFS is then ideally suited for Extreme-AO:
\begin{itemize}
\item{A very fast linear wavefront reconstruction can be employed}
\item{The nlCWFS is achromatic and can work in broadband}
\item{The nlCWFS can measure with full sensitivity wavefronts which are far from flat. The nlCWFS can be operated in a linear control loop around a "reference" wavefront which is not flat. This feature is very useful when non-common path errors need to be removed for example.}
\end{itemize}
Thanks to the nlCWFS linearity, computing requirements can be greatly relaxed for ExAO: a single matrix multiplication would be performed to reconstruct the wavefront. The linearity relationship also facilitates spreading the computing load between CPUs, which can be done more easily than for the iterative algorithm proposed in Figure \ref{fig:wfsalgo}.

\subsubsection{Expected performance in a coronagraphic Extreme-AO system}

The performance of a nlCWFS-based extreme-AO system can be quantified using the same formalism as in \citet{guyo05}. In this section, only the PSF halo contributions due to time lag in the AO loop and photon noise in the WFS are considered (term $C_2$ in \citet{guyo05}). Scintillation and optical pathlength chromaticity between sensing and imaging wavelength are not taken into account.

\begin{figure*}[tbh]
\includegraphics[scale=0.32]{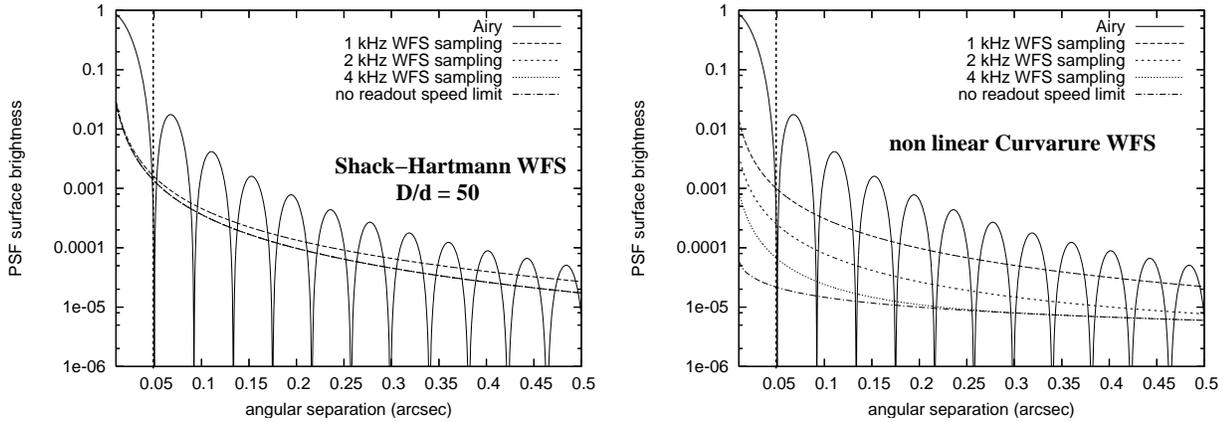}
\caption{\label{fig:contvsspeed} Achievable PSF contrast $C_2$ (y-axis) as a function of angular separation (x-axis) for an AO system measuring wavefront aberrations at $\lambda=0.7 \mu m$ with $7.8\:10^7$ ph.s$^{-1}$ and acquiring scientific images at $\lambda_i=1.6 \mu m$ on a 8-m diameter telescope. The PSF surface brightness is shown for several WFS frame readout frequencies for both a Shack-Hartmann WFS (left) and a non linear Curvature WFS (right).}
\end{figure*}


Figure \ref{fig:contvsspeed} shows, for a photon rate of $7.8\:10^7$ ph.s$^{-1}$ in the WFS ($m_V\approx7$ star observed in a 0.1 $\mu$m wide band with no loss in the optics and an ideal detector), how the PSF halo surface brightness $C_2$ varies with both spatial frequency and WFS sampling time. With a SHWFS with $D/d=50$, 
PSF halo surface brightness is dominated by photon noise in the WFS for sampling rates beyond 1 kHz (Figure \ref{fig:contvsspeed}): increasing the WFS sampling rate beyond 1 kHz offers no benefit. The nlCWFS's high sensitivity allows performance (lower PSF halo surface brightness) with higher WFS sampling rates. The performance difference is especially large at small angular separation, where a full order of magnitude can be gained if the WFS can be sufficiently fast.

\section{Conclusions}
Simulations and analysis presented in this paper show that the non linear Curvature WFS technique is potentially able to work close to the fundamental theoretical sensitivity limit imposed by photon noise even if the PSF is not diffraction limited at the WFS wavelength. It is also a highly flexible WFS choice, as it can efficiently sense wavefront on sparse pupils, and performs an explicit wavefront reconstruction, therefore allowing open-loop operation and compensation of non-common path errors.

In the preliminary study presented in this paper, encouraging results were obtained on intermediate brightness guide stars with closed loop simulations using a relatively simple control algorithm. Further performance gains are to be expected if both modal control (optimal weighting of wavefront modes according to their measurement SNR and variance in the atmospheric turbulence) and predictive control are also included. A key challenge to fully take advantage of this technique is to achieve sufficiently high AO loop frequency, and algorithms with faster convergence than the iterative loop used in this paper should be developed.

The nlCWFS appears especially well suited for Extreme-AO systems aimed at delivering nearly flat wavefronts to a high contrast imaging camera (coronagraph) for direct imaging of exoplanets and disks. In these systems a nlCWFS can deliver for low order aberrations on a 8m telescope the same wavefront measurement accuracy as a SHWFS with $\approx 1000$ times fewer photons. The nlCWFS can take full advantage of the telescope's diffraction limit for wavefront sensing, with a sensitivity scaling as $D^4$ (combination of a $D^2$ gain due smaller diffraction limit and a $D^2$ gain due to the telescope's collecting power) while more conventional WFSs scale only as $D^2$ (telescope's collecting power). This difference is especially large for the next generation of Extremely Large Telescopes (ELTs), where the use of high sensitivity WFS techniques such as the nlCWFS may allow direct imaging of exoplanets in reflected light.

\acknowledgements
The author is thankful to Roger Angel, Phil Hinz, Michael Hart, Markus Kasper, Marcos van Dam and Christophe Verinaud for valuable comments on this manuscript.

\end{document}